%% file: ms.tex
\documentclass[a4paper,11pt]{article}
\pdfoutput=1 % if your are submitting a pdflatex (i.e. if you have
\usepackage{jcappub}
\usepackage{graphicx}
\usepackage{color}
\usepackage{makecell}
\usepackage{afterpage}
\usepackage[dvipsnames]{xcolor}
\usepackage{multirow}
\usepackage{cleveref}
\usepackage[normalem]{ulem}		% Automatic eq. fig., etc.
\usepackage{xspace}
\usepackage{subcaption}
\usepackage{booktabs}

\usepackage{centernot}
\usepackage{mathtools}
\usepackage{stmaryrd}
\usepackage[utf8]{inputenc}

\usepackage{placeins}
\usepackage{titlesec}
\titlespacing*{\section} {0pt}{1.5ex plus 0.5ex minus .2ex}{-1.8ex plus .2ex}
\titlespacing*{\subsection} {0pt}{0.75ex plus 0.5ex minus .2ex}{-1.8ex plus .2ex}

\newcommand{\tquote}[1]{``#1''}

\newcolumntype{L}[1]{>{\raggedright\arraybackslash}p{#1}}
\newcolumntype{C}[1]{>{\centering\arraybackslash}p{#1}}

\newcommand{\class}{{\sc class}\xspace}

\newcommand{\omegab}{\omega_\mathrm{b}}

\newcommand{\Omegam}{\Omega_\mathrm{m}}
\newcommand{\Neff}{N_\mathrm{eff}}

\title{The BAO+BBN take on the Hubble tension}

\author[1]{Nils Sch\"oneberg,}
\author[1]{Julien Lesgourgues,} 
\author[1]{and Deanna C. Hooper} 

\affiliation[1]{Institute for Theoretical Particle Physics and Cosmology (TTK), \\ RWTH Aachen University, D-52056 Aachen, Germany.}

\emailAdd{schoeneberg@physik.rwth-aachen.de}
\emailAdd{lesgourg@physik.rwth-aachen.de}
\emailAdd{hooper@physik.rwth-aachen.de}

\abstract{Many attempts to solve the Hubble tension with extended cosmological models combine an enhanced relic radiation density, acting at the level of background cosmology, with new physical ingredients affecting the evolution of cosmological perturbations. Several authors have pointed out the ability of combined Baryon Acoustic Oscillation (BAO) and Big Bang Nucleosynthesis (BBN) data to probe the background cosmological history independently of both CMB maps and supernovae data. 
Using state-of-the-art assumptions on BBN, we confirm that combined BAO, deuterium, and helium data are in tension with the SH0ES measurements under the $\Lambda$CDM assumption at the 3.2$\sigma$ level, while being in close agreement with the CMB value. We subsequently show that floating the radiation density parameter $N_{\rm eff} $ only reduces the tension down to the 2.6$\sigma$ level. This conclusion, totally independent of any CMB data, shows that a high $N_\mathrm{eff}$ accounting for extra relics (either free-streaming  or self-interacting) does not provide an obvious solution to the crisis, not even at the level of background cosmology. To circumvent this strong bound, (i) the extra radiation has to be generated after BBN to avoid helium bounds, and (ii) additional ingredients have to be invoked at the level of perturbations to reconcile this extra radiation with CMB and LSS data.}

\begin{document}

\hfill{\small TTK-19-30}

\maketitle

\setlength{\parskip}{\baselineskip}%
\input{Introduction.tex}
\input{Data.tex}
\input{Results.tex}
\input{Discussion.tex}

\section*{Acknowledgements}
We thank the authors of \cite{Cuceu:2019for} for useful discussions and providing their Lyman-$\alpha$ likelihood for comparison. NS acknowledges support from the DFG grant LE 3742/4-1. DH and JL are supported by the DFG grant LE 3742/3-1. Simulations were performed with computing resources granted by RWTH Aachen University under project jara0184.

%\clearpage
\appendix
\input{BAO.tex}

%\clearpage

\bibliography{biblio}{}
\bibliographystyle{JHEP}

\end{document}

%% file: Introduction.tex
\section{Introduction}
\label{sec:intro}
{\it Tensions in cosmological data.} CMB data from the Planck satellite~\cite{Aghanim:2018eyx} offer a very consistent picture of the cosmological model, with internal consistency between the parameters extracted from the four independent spectra $C_\ell^{TT}$, $C_\ell^{TE}$, $C_\ell^{EE}$, $C_\ell^{\phi \phi}$. Furthermore, there is excellent agreement with other CMB observations, with measurements of the Baryon Acoustic Oscillation (BAO) scale, with the abundance of primordial elements produced by Big Bang Nucleosynthesis (BBN), with high-redshift supernovae probing the acceleration of the Universe expansion, with the overall shape of the galaxy power spectrum, and with the measurement of redshift space distortions in galaxy surveys. However, the Planck best-fit model is in strong $4.4\sigma$ tension with direct measurements of the current expansion rate $H_0$ obtained by the SH0ES collaboration using nearby supernovae calibrated to Cepheids~\cite{Riess:2019cxk}, and in mild tension with other $H_0$ measurements based on quasar time delays~\cite{Wong:2019kwg} (while nearby supernovae calibrated through the Tip of the Red Giant Branch currently agree with both Planck and SH0ES values \cite{Freedman:2019jwv}).  A mild tension is also reported with most cosmic shear surveys and cluster count experiments regarding the value of the parameter $\sigma_8$~(see e.g. \cite{Joudaki:2019pmv} and references therein).

These tensions are often described as an inconsistency between probes of the early universe and of the late universe, but this is somewhat oversimplified, since both CMB and BAO data mix information coming from the early and late cosmological evolution in a subtle way (these two types of information can be disentangled with dedicated approaches~\cite{Vonlanthen:2010cd,Audren:2012wb,Audren:2013nwa}, but at the cost of losing constraining power). 

Interestingly, these tensions appear in the determination of two parameters, $H_0$ and $\sigma_8$, that are probed by CMB data in a very indirect -- and thus model-dependent -- way. Extended cosmological models can provide as good a fit to CMB data as the minimal $\Lambda$CDM model, while still leading to different predictions for $H_0$ and $\sigma_8$. Therefore, there is some hope that the apparent tensions result from assuming the wrong cosmological model, and many cosmologists are investigating scenarios that could reconcile all data sets. In practice, this turns out to be difficult, as simple extensions of $\Lambda$CDM fail to restore agreement. The Hubble tension -- and potentially also the $\sigma_8$ tension -- are only reduced with rather non-trivial assumptions, like self-interacting active neutrinos plus extra radiation~\cite{Lancaster:2017ksf,Oldengott:2017fhy,DiValentino:2017oaw,Kreisch:2019yzn,Park:2019ibn}, a light sterile neutrino interacting with a scalar field~\cite{Archidiacono:2016kkh}, Dark Matter and Dark Radiation particles belonging to the same Dark Sector and interacting with each other~\cite{Lesgourgues:2015wza,Buen-Abad:2017gxg,Archidiacono:2019wdp}, Dark Matter converting into Dark Radiation~\cite{Poulin:2016nat,Binder:2017lkj,Bringmann:2018jpr}, Dark Radiation from PBH~\cite{Hooper:2019gtx}, early Dark Energy~\cite{Poulin:2018cxd,Agrawal:2019lmo,Lin:2019qug}, 
Dark Matter interacting with Dark Energy~(\cite{Pan:2019gop} and references therein), fifth force effects \cite{Desmond:2019ygn},
and many others. Many of the models quoted here (\cite{Lancaster:2017ksf,Oldengott:2017fhy,DiValentino:2017oaw,Kreisch:2019yzn,Park:2019ibn,Archidiacono:2016kkh,Lesgourgues:2015wza,Buen-Abad:2017gxg,Archidiacono:2019wdp,Poulin:2016nat,Binder:2017lkj,Bringmann:2018jpr,Hooper:2019gtx} plus sub-cases of \cite{Poulin:2018cxd}) share a common ingredient: an enhanced radiation density, at least until the time of photon-baryon decoupling. We will come back to this aspect which is related to the main point of this work.

\noindent {\it The BAO+BBN probe.} Since the debate on the Hubble tension is often presented as a direct confrontation between CMB data and direct $H_0$ measurements, it is interesting to devise independent approaches providing estimates of $H_0$. Several authors \cite{Addison:2013haa,Aubourg:2014yra,Addison:2017fdm,Blomqvist:2019rah,Cuceu:2019for} have proposed such an approached based on combined BAO and BBN data.

In the $\Lambda$CDM framework, BAO data provide confidence limits in the three-dimensional parameter space $(\Omegam, H_0, \omegab)$. BAO data alone do not provide accurate constraints on $H_0$ after marginalizing over $(\Omegam, \omegab)$. 
To obtain such constraints, an independent measurement of $\omegab$ is needed. While this could be done by using the Planck predictions for $\omegab$, the results would not be independent of CMB data. Instead, we can use measurements of the primordial deuterium abundance, which, combined with standard BBN assumptions, provide a robust measurement of $\omegab$. Blomqvist et al.~\cite{Blomqvist:2019rah} and Cuceu et al.~\cite{Cuceu:2019for} showed that joint BAO+deuterium data lead to narrow contours in the $(H_0, \Omegam)$ plane, and to interesting marginalized bounds on $H_0$ alone. This new result is not just driven by the idea to combine BAO and BBN data, but also by the role of recent BAO measurements inferred from \mbox{Lyman-$\alpha$} data in the Data Release 14 (DR14) of the eBOSS survey~\cite{Agathe:2019vsu,Blomqvist:2019rah}. Previous data releases showed a tension between the BAO scale inferred from the BOSS galaxy survey and Lyman-$\alpha$ survey. However, in DR14, a substantial amount of additional data has reduced the tension between these surveys \cite{Agathe:2019vsu,Blomqvist:2019rah},  such that combining all BAO measurements makes sense statistically, as recently pointed out by Cuceu et al. \cite{Cuceu:2019for}. The overlap between the BAO-galaxy + BBN and the BAO-Lyman-$\alpha$ + BBN preferred regions is small, and favors low values of $H_0$ in better agreement with Planck than with SH0ES \cite{Cuceu:2019for}. 

The combined BAO+BBN probe is particularly interesting because it is only sensitive to the evolution of background quantities. Indeed, primordial element abundances, cosmological distances, and the BAO scale\footnote{The BAO scale could be seen as a joint probe of the background evolution, through the sound horizon that depends only on background quantities, and of the perturbation evolution, through effects like \tquote{neutrino drag} \cite{Bashinsky:2003tk,Hou:2011ec,Lesgourgues:2018ncw}. In Appendix A we show that the latter effects are negligible given the current level of precision of BAO data. This conclusion is consistent with an earlier analysis by Thepsuriya and Lewis~\cite{Thepsuriya:2014zda}. Thus, the BAO scale essentially probes the sound horizon and the average pressure of the baryon-photon fluid before the baryon drag time~\cite{Eisenstein:1997ik}.} only depend on the assumed background cosmology. Many models that have been proposed to solve the $H_0$ and $\sigma_8$ tensions differ from each other at the level of cosmological perturbations, but are degenerate at the background level. Indeed, the background history is independent of several ingredients in these models, like e.g. scattering or self-interactions in the sector of neutrinos, Dark Matter, Dark Radiation, or Dark Energy. It only probes the homogeneous density of these new species, captured by $N_\mathrm{eff}$ for any relativistic relics, $\omega_\mathrm{cdm}$ for cold dark matter, and a combination of the two for hot or warm dark matter. Hence, the BAO+BBN probe yields constraints applying to many extended models at once.

\noindent {\it The $H_0 -  \Neff$ degeneracy.} Many of the extended models attempting to solve the Hubble tension feature additional relativistic relics, at least until the time of radiation/matter equality, and thus an enhanced value of $\Neff$. Indeed, increasing $\Neff$ is a rather generic way to reduce the tension between CMB and SH0ES measurements. The reason is simply that the CMB is extremely sensitive to ratios between background densities: the ratio of radiation to matter, or that of matter to $\Lambda$. Increasing $H_0$ means increasing the critical density today, and thus the density of matter and $\Lambda$. This can be done with a fixed  matter-to-$\Lambda$ ratio. The  radiation-to-matter ratio can also be kept fixed if $\Neff$ increases together with $H_0$. This leads to the famous $H_0 -  \Neff$ degeneracy \cite{Bashinsky:2003tk,Hou:2011ec,Lesgourgues:2018ncw}. Nevertheless, this degeneracy is broken by CMB data, which are sensitive to other consequences of this transformation (like extra Silk damping, or an increased CMB peak shift due to \tquote{neutrino drag} \cite{Bashinsky:2003tk,Hou:2011ec,Lesgourgues:2018ncw}). The art of solving the Hubble tension consists in introducing new ingredients in the cosmological model -- usually at the level of perturbations -- that will counteract these other consequences. Only then, both $H_0$ and  $\Neff$ can be increased significantly without deforming the CMB and matter power spectra too much. This is how self-interactions or scattering processes in the sector of neutrinos, Dark Matter, Dark Radiation, or Dark Energy are called to the rescue. Note that there exist other categories of model aimed at solving the Hubble tension with very different ingredients and no enhancement of $\Neff$ (see e.g.~\cite{Pan:2019gop,Desmond:2019ygn} and references therein).

\noindent {\it Goal.} Here we study to which extent the BAO+BBN probe constrains the $H_0 -  \Neff$ degeneracy at the level of background cosmology. Addressing this point is especially interesting since the BAO+BBN bounds will apply to several models mentioned previously, which are degenerate with the $\Lambda$CDM+$\Neff$ model at the background level. The goal of our analysis is not trivial because varying $\Neff$ has several consequences for the quantities probed by BAO+BBN data. First, $\Neff$ affects the predictions of BBN, such as the primordial deuterium and helium abundance. Thus, we will use joint deuterium and helium measurements to constrain simultaneously $\omegab$ and $\Neff$\,, still independently of any CMB data. Second, $\Neff$ affects the redshift of equality between matter and radiation and thus the sound horizon. These two effects are intricate, and a dedicated analysis is needed to check whether floating $\Neff$ may reduce the tension between the value of $H_0$ inferred from the BAO+BBN probe and from SH0ES.

In section~\ref{sec:data} we will introduce our BAO and BBN data sets, as well as the direct $H_0$ measurements and CMB data that we use for comparison. In the results section~\ref{sec:results}, we will first attempt to confirm the results of Cuceu et al.~\cite{Cuceu:2019for} assuming a minimal $\Lambda$CDM cosmology. However, we will use both primordial deuterium and helium measurements, and we will rely on an improved treatment of primordial abundances -- in particular, a well-motivated theoretical error on BBN predictions. Then we will extend our analysis to the case of a free $\Neff$ in subsection~\ref{subsec:neff}. Our conclusions are presented in~section~\ref{sec:discussion}.

%% file: Data.tex
\newpage
\section{Data and assumptions}
\label{sec:data}

\subsection{BAO}

\begin{table}[t]
%\hspace*{-5mm}
\begin{tabular}{| c | cc c |}
         \hline
	Measurement & $z_\mathrm{BAO}$ & Value & Ref. \\ \hline
	\textbf{BAO from galactic data} & & & \\
	6DF & 0.106 & $r_s/D_V = 0.327\pm0.015$ & \cite{2011MNRAS.416.3017B}\\
	SDSS DR7 MGS & 0.15 & $D_V/r_s = 4.47\pm0.16$ & \cite{Ross:2014qpa} \\ 
	%SDSS DR14 eBOSS LRG & 0.72 & $D_V/r_s =15.92\pm0.42$ & \cite{Bautista:2017wwp} \\
	SDSS DR14 eBOSS QSO & 1.52 & $D_V/r_s =26.0\pm 1.0$ & \cite{Ata:2017dya}\\
	SDSS DR12 galaxies &  0.38, 0.51, 0.61& full covariance matrix & \cite{Alam:2016hwk} \\
	\hline
	\textbf{BAO from Lyman-$\boldsymbol{\alpha}$ data} & & & \\
	SDSS DR14 eBOSS Ly-$\alpha$ & 2.34 &$1/(H r_s) = 8.86\pm0.29$ & \cite{Agathe:2019vsu}\\
	SDSS DR14 eBOSS Ly-$\alpha$ & 2.34 &$D_A/r_s = 11.20\pm0.56$ & \cite{Agathe:2019vsu}\\
	SDSS DR14 eBOSS Ly-$\alpha$-QSO cross & 2.35 &$1/(H r_s) = 9.20\pm0.36$ & \cite{Blomqvist:2019rah} \\
	SDSS DR14 eBOSS Ly-$\alpha$-QSO cross & 2.35 &$D_A/r_s = 10.84\pm0.54$ & \cite{Blomqvist:2019rah}\\
	\hline
\end{tabular}
	\caption{List of BAO measurements used in this work.\label{tab:bao_measures}}
\end{table}

We updated the BAO likelihood in the {\sc MontePython}\footnote{\url{https://github.com/brinckmann/montepython_public}}~\cite{Audren:2012wb,Brinckmann:2018cvx} sampler in order to include all the BAO scale measurements listed in table~\ref{tab:bao_measures}. They are divided in two categories: the measurements inferred from galaxy redshift surveys, and those extracted from  Lyman-$\alpha$ data. The experimental collaborations measure three types of ratios of comoving distances: the angular scale of the BAO ($D_A/r_s$), the redshift-space BAO scale ($1/H r_s$), and the spherically-averaged BAO scale ($D_V/r_s$). For each point, we use a simple Gaussian likelihood comparing the measured ratio with the \class\footnote{\url{http://class-code.net}}~\cite{Blas:2011rf} predictions. \class computes $r_s$, $D_A$, $H$, and $D_V$ with high accuracy. The relative difference between these numbers computed either by {\sc camb}~\cite{Lewis:1999bs} or \class is at most of the order of $10^{-5}$, while the most precise measurement (angular BAO scales in SDSS DR12) have a $\sim 1\%$ error. At this level of accuracy, modeling the measurement of the BAO scale as a measurement of the sound horizon is an excellent approximation, even when one varies $\Neff$ and the associated \tquote{neutrino drag} effect (see ref.~\cite{Thepsuriya:2014zda} and Appendix  \ref{ap:bao_measure} for more details).

\subsection{BBN}
We implemented the BBN likelihoods in {\sc MontePython} which share exactly the same assumptions as in the BBN section of the Planck 2018 cosmological parameter paper~\cite{Aghanim:2018eyx}. BBN data consist of measurements of the primordial abundances of helium, $Y_P$, and deuterium, $y_\mathrm{DP}=10^5 n_\mathrm{D}/n_\mathrm{H}$. For each of them we use a Gaussian likelihood:
\begin{equation}
\ln {\cal L} = - \frac{1}{2} \frac{(Y_\mathrm{obs} - Y_\mathrm{th} (\omegab, \Neff))^2}{\sigma_\mathrm{obs}^2 +  \sigma_\mathrm{th}^2}~,
\end{equation}
where $Y$ stands for either $Y_P$ or $y_\mathrm{DP}$,  $Y_\mathrm{obs}$ and $\sigma_\mathrm{obs}$ are the measured value and measurement error, $Y_\mathrm{th} (\omegab, \Neff)$ is the theoretical prediction based on a BBN code, and $\sigma_\mathrm{th}$ is the theoretical error propagating all uncertainties on assumptions done in the code, like the neutron lifetime, nuclear reaction rates, etc.

We use the latest deuterium measurement of Cooke et al. \cite{Cooke:2017cwo}, $y_\mathrm{DP} = 2.527\pm0.030$ (68\%C.L.). For helium, our baseline is the latest helium data compilation of Aver et al. \cite{Aver:2015iza}, $Y_P = 0.2449 \pm 0.0040$ (68\%C.L.), but we also check the impact of using other estimates by Peimbert et al. \cite{Peimbert:2016bdg} and Izotov et al. \cite{Izotov:2014fga}.

For theoretical predictions, our baseline likelihood uses the code {\tt PArthENoPE 2.0}~\cite{Consiglio:2017pot} with an estimate of nuclear rate $d(p,\gamma)^3$He derived from the observations of Adelberger et al. \cite{Adelberger:2010qa}. Instead of running {\tt PArthENoPE} on-the-fly during each {\sc MontePyhton} run, we produced some interpolation tables in advance for the helium and deuterium abundance as a function of $(\omegab, \Neff)$, which are stored within the likelihood. Following the authors of {\tt PArthENoPE}, for helium we assume a theoretical error $\sigma_\mathrm{th} = 3.0 \times 10^{-4}$ (arising mainly from the uncertainty on the neutron lifetime), while for deuterium we take $\sigma_\mathrm{th} = 0.060$ (arising from uncertainties on several nuclear rates). We call this baseline BBN model \tquote{{\tt PArthENoPE}-standard}. 

To assess the robustness of our results we will also employ two other BBN models. The second one, called \tquote{{\tt PArthENoPE}-Marcucci}, also relies on {\tt PArthENoPE 2.0}, but takes the nuclear rate $d(p,\gamma)^3$He from the {\it ab initio} calculation of Marcucci et al.~\cite{Marcucci:2015yla}, and consistently reduces the theoretical error on deuterium to $\sigma_\mathrm{th} = 0.030$. The third one, called \tquote{{\tt PRIMAT}}, uses the BBN code of Pitrou et al.~\cite{Pitrou:2018cgg}, in which the nuclear rate $d(p,\gamma)^3$He takes into account both laboratory experiments and {\it ab initio} calculations. In this case, following the authors of {\tt PRIMAT}, we reduce the theoretical error on deuterium to $\sigma_\mathrm{th} = 0.032$.

\begin{figure}
	\includegraphics[width=\textwidth]{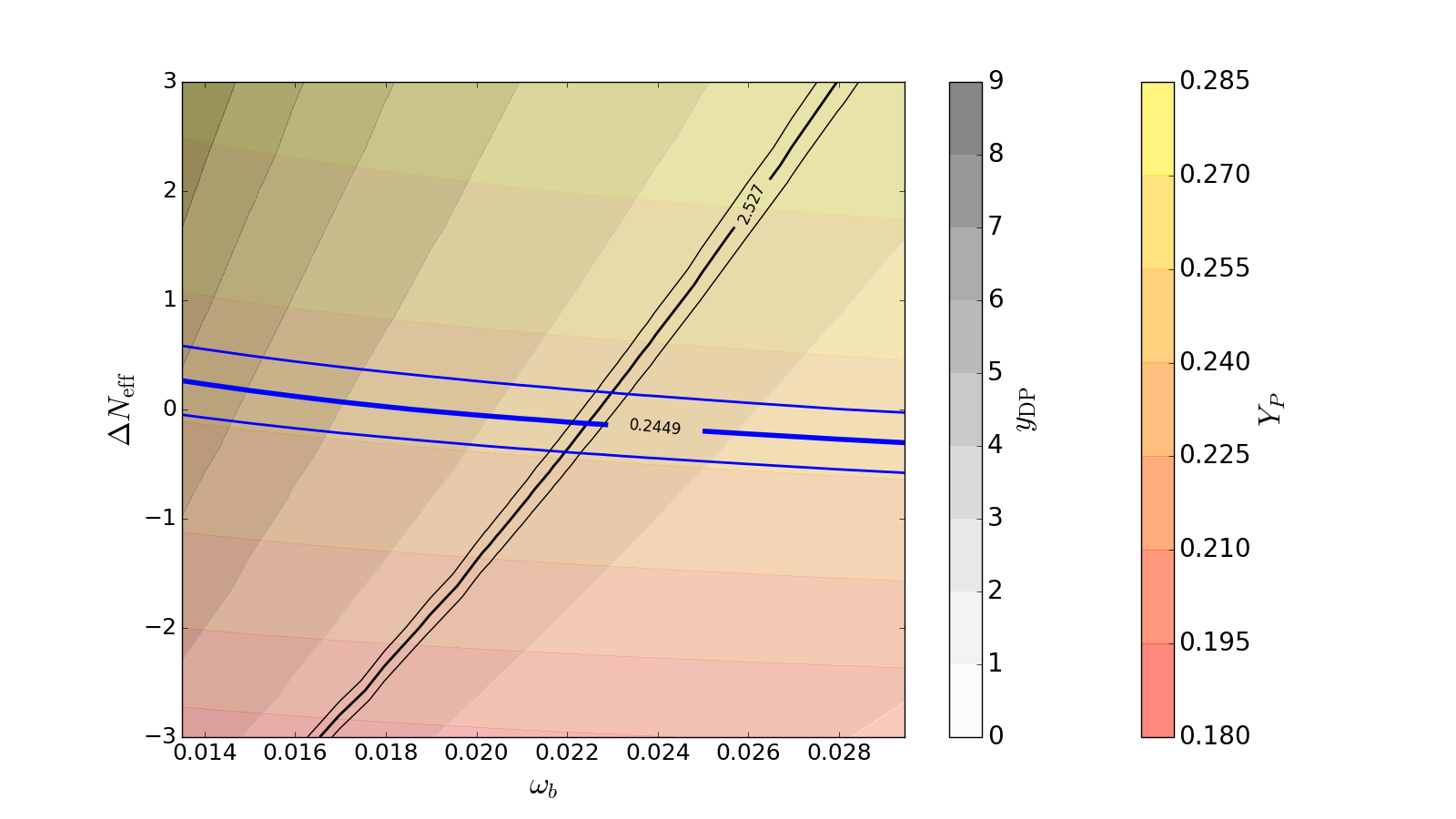}
	\caption{Primordial abundances $Y_P$ (colored) and $y_\mathrm{DP}$ (grey) as a function of $(\omegab, \Delta \Neff)$ with the {\tt PArthENoPE}-standard BBN model, with superimposed measurements of  Aver et al. \cite{Aver:2015iza} for $Y_P$ (blue) and Cooke et al.  \cite{Cooke:2017cwo} for  $y_\mathrm{DP}$ (black). We can see that a $\Delta N_\mathrm{eff} = N_\mathrm{eff}-3.046$ around zero is preferred, and deviations from it are mostly constrained by the $Y_P$ measurement.\label{fig:BBN}}
\end{figure}

Our BBN treatment improves over previous studies of combined BAO+BBN data~\mbox{\cite{Blomqvist:2019rah,Cuceu:2019for}}, where a Gaussian prior on $\omegab$ based on the deuterium  measurement of Cooke et al. \cite{Cooke:2017cwo} was used. Indeed, assuming $\Neff=3.046$, Cooke et al. converted their deuterium data point into estimates of $\omegab$ using the BBN code of Nollett et al. \cite{Nollett:2011aa}, which makes different assumptions on the deuterium fusion rates $d(d, n)^3$He and $d(d, p)^3$H than PArthENoPE or PRIMAT. The code of Nollett et al. predicts a smaller deuterium abundance than PArthENoPE by about $\Delta y_\mathrm{DP}=0.04$, and assesses a smaller theoretical error. The most recent versions of PArthENoPE or PRIMAT predict values of $y_\mathrm{DP}$  close to each other by about $\Delta y_\mathrm{DP}=0.015$, and assess conservative theoretical errors, of the same order of magnitude as the shifts between the three codes. Thus, our pipeline provides more conservative constraints on $\omegab$ (and also $\Neff$). Figure~\ref{fig:BBN} illustrates the  constraining power of our baseline BBN likelihood.

\subsection{Comparative measurements}

We will compare our BAO+BBN results with two completely independent data sets. 

The first one is the direct $H_0$ measurement recently inferred from supernovae calibrated on Cepheids by the SH0ES collaboration, who obtained $H_0 = 74.03 \pm 1.42$ km s$^{-1}$ Mpc$^{-1}$(68\%C.L.) \cite{Riess:2019cxk}. Since the purpose of this work is to discuss the $H_0$ tension, we base our comparisons on the highest measured value of $H_0$: recent data from the Carnegie-Chicago~\cite{Freedman:2019jwv} or H0LiCOW~\cite{Wong:2019kwg} collaborations are compatible with smaller values of $H_0$.

The second one is CMB data from the Planck 2018 release, including high-$\ell$ TTTEEE and low-$\ell$ temperature and polarization data. The Planck contours presented in this work  are taken directly from the chains available on the Planck Legacy Archive\footnote{\url{https://pla.esac.esa.int}}.  

%% file: Results.tex
\section{Results}\label{sec:results}

We ran {\sc class}+{\sc MontePython} in adaptive Metropolis-Hastings mode~\cite{Brinckmann:2018cvx} to derive constraints on cosmological parameters from the BAO+BBN probe, first for the minimal $\Lambda$CDM model (section~\ref{sec:vanilla}), and then for the $\Lambda$CDM+$\Neff$ model (section~\ref{subsec:neff}). All of our runs reached a Gelman-Rubin convergence criterion of $|R-1|<10^{-4}$.

\subsection{$\Lambda$CDM model} \label{sec:vanilla}

\begin{figure}[t!]
	\includegraphics{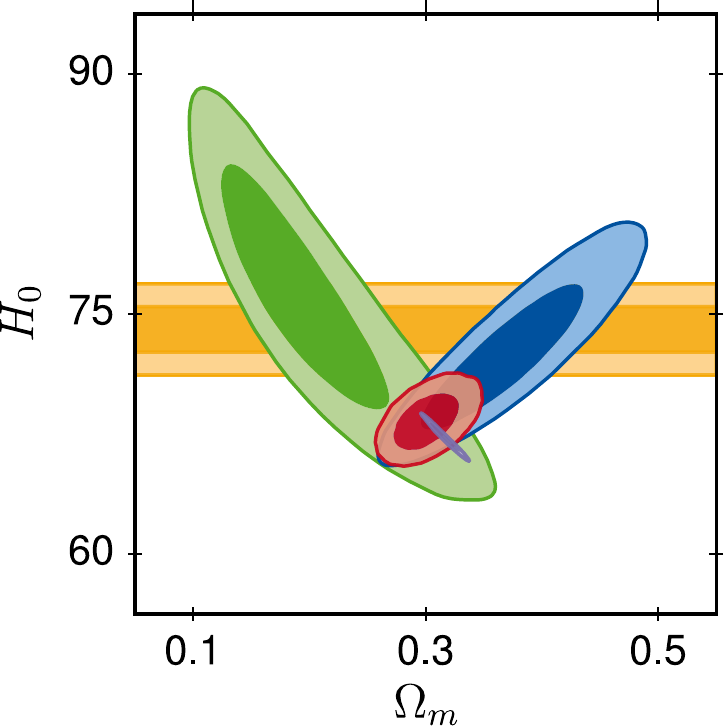}
	\includegraphics{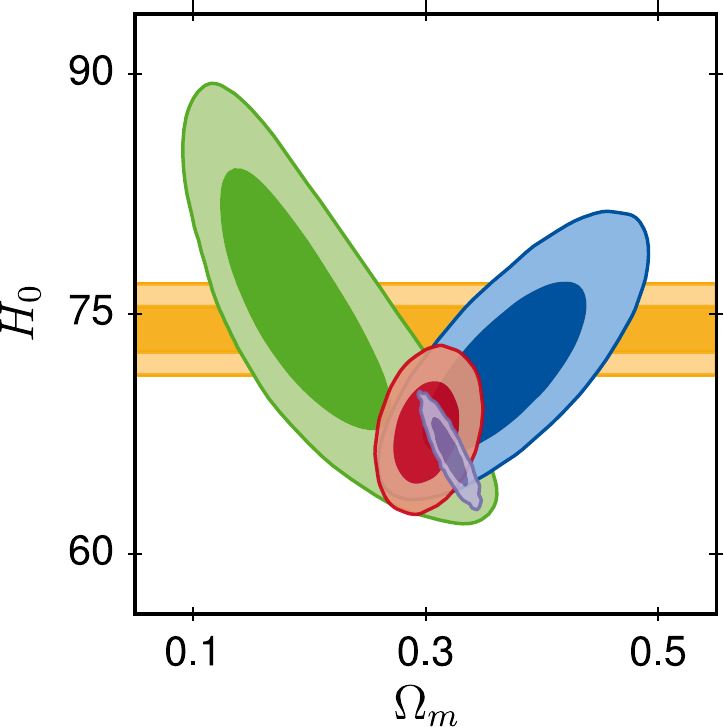}
	\caption{68\% and 95\% confidence levels on $\Omega_m$ and $H_0$ with various data sets: BAO+BBN for galaxy BAO (blue), Lyman-$\alpha$ BAO (green), and combined (red). Additionally displayed is the Riess et al 2019 measurement (orange), and the Planck 2018 measurement (purple). \textbf{Left:} Minimal $\Lambda$CDM model. \textbf{Right:} The $\Lambda$CDM model extended with $N_\mathrm{eff}$\,.
	\label{fig:lyagal_nonur}}
\end{figure}

For the minimal $\Lambda$CDM model, we adopt flat priors on the three parameters $(\Omegam, H_0, \omegab)$. The other $\Lambda$CDM parameters are not probed by BAO+BBN experiments and are thus kept fixed. As in the Planck baseline model, we assume one massive neutrino species with $m=0.06\,$eV, and two massless ones.

In order to test our pipeline, we first run with the same assumptions as Cuceu et al.~\cite{Cuceu:2019for}: we replace our full BBN likelihoods with a Gaussian prior on $\omegab$ taken from Cooke et al. \cite{Cooke:2017cwo} (for the case with a theoretical determination of the nuclear rate $d(p,\gamma)^3$He). The only remaining difference with reference \cite{Cuceu:2019for} is the calculation of the comoving sound horizon: our BAO likelihood uses the value calculated by \class instead of an analytic approximation. Nevertheless, our results agree very well with \cite{Cuceu:2019for} (and also \cite{Blomqvist:2019rah}), with a final prediction of $H_0 = 67.5 \pm 1.1$\,km/s/Mpc (68\%C.L.) for the combined data set, in 3.6$\sigma$ tension with the SH0ES measurement. 

We then switch to our own assumptions described in section~\ref{sec:data}: we use the full BAO data described in table~\ref{tab:bao_measures} plus our BBN likelihoods for both deuterium and helium, with the baseline combination ({\tt PArthENoPE}-standard and helium data from Aver et al. \cite{Aver:2015iza}). Our results are shown in figure~\ref{fig:lyagal_nonur}. We first run the two BAO data subsets individually, to illustrate their different degeneracy directions, and then their combination to show the full constraining power. With the full BAO+BBN data we obtain the marginalized bounds $H_0 = 68.3 \substack{+1.1 \\ -1.2}$\,km/s/Mpc (68\%C.L.), at 3.2$\sigma$ tension with the SH0ES collaboration measurement, but within 0.7$\sigma$ of the Planck 2018 measurement ($H_0 = 67.3\pm0.6\,$km/s/Mpc, 68\%C.L.). Our results are slightly less in tension with SH0ES than those of \cite{Cuceu:2019for,Blomqvist:2019rah} because we use a realistic theoretical error on deuterium abundance predictions. We still reach the same qualitative conclusion as these references: in the $\Lambda$CDM framework, BAO data are in tension with direct measurement of the Hubble parameter, even when taking a conservative prior on $\omegab$ independent of any CMB data.

\begin{figure}[t!]
\includegraphics{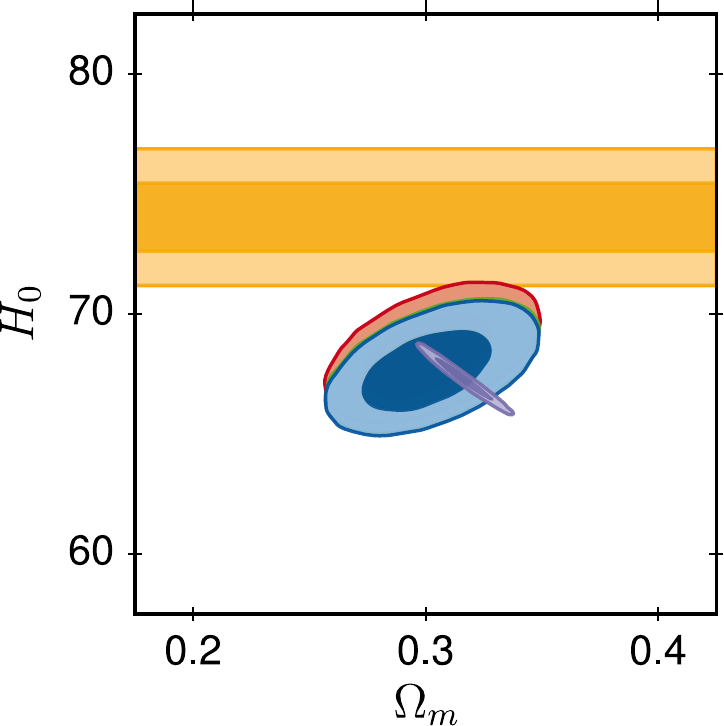}
\includegraphics{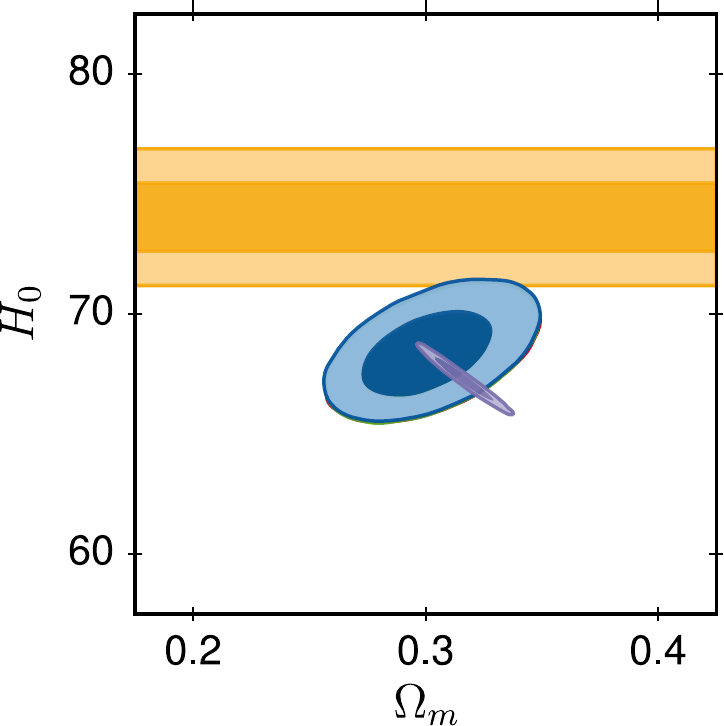}
\caption{68\% and 95\% confidence levels on $\Omega_m$ and $H_0$ for the minimal $\Lambda$CDM model and our combined BAO+BBN data set under various assumptions: \textbf{Left:} BBN predictions taken from {\tt PArthENoPE}-standard (red), {\tt PArthENoPE}-Marcucci (green), and {\tt PRIMAT} (blue). \textbf{Right:} Helium abundance taken from Aver et al. \cite{Aver:2015iza} (red), Peimbert et al. \cite{Peimbert:2016bdg} (green), and Izotov et al. \cite{Izotov:2014fga} (blue).
\label{fig:bbn_tests_nonur}}
\end{figure}

We performed additional runs to test the influence of different BBN modeling and helium measurements on this result. Our findings are shown in figure \ref{fig:bbn_tests_nonur}. The left panel shows the impact of switching between the three pipelines {\tt PArthENoPE}-standard, {\tt PArthENoPE}-Marcucci and {\tt PRIMAT}.  Essentially, this amounts in taking different assumptions on the most important nuclear rates. Cuceu et al. \cite{Cuceu:2019for}, performed a similar test and found that their results for $H_0$ vary by 0.45$\sigma$ depending on such assumptions. In our case the variations remains within less than 0.1$\sigma$ because in each case the theoretical error already takes into account the uncertainties on these rates. The right panel shows the impact of taking the experimental helium abundance measurement from the three different references listed in section~\ref{sec:data}. This has a totally negligible impact because $\omegab$ depends on deuterium much more than helium measurements. In conclusion, we find that our results are robust and hardly affected by the most controversial aspects of BBN physics.

\subsection{$\Lambda$CDM + $N_\mathrm{eff}$ model}
\label{subsec:neff}

We now allow $N_\mathrm{eff}$ to vary in order to check whether the BAO+BBN dataset is compatible with larger values of $H_0$. Different physical effects are now at play. 

BAO data probe the ratio of the comoving sound horizon over several types of comoving cosmological distances. For instance, in a flat universe, the BAO angular scale reads:
\begin{equation}
\theta(z) = \frac{\int_{z_D}^\infty c_s(\omegab, \tilde{z}) H(\tilde{z})^{-1} d\tilde{z}}{\int_0^z H(\tilde{z})^{-1} d\tilde{z}}
\simeq \frac{\int_{z_D}^\infty c_s(\omegab, \tilde{z}) \left[\Omega_\mathrm{r}/\Omegam (1+\tilde{z})^4 + (1+\tilde{z})^3\right]^{-1/2} d\tilde{z}}{\int_0^z \left[1/\Omegam-1 + (1+\tilde{z})^3\right]^{-1/2} d\tilde{z}}~,
\end{equation}
where the baryon-photon sound speed $c_s$ depends on the baryon density and on redshift.
For a fixed $\omegab$, the numerator (related to the sound horizon) depends on the redshift of equality between radiation and matter (i.e. on $\Omegam/\Omega_\mathrm{r}$), while the denominator (related to the angular diameter distance) depends on the redshift of equality between matter and $\Lambda$ (i.e. on $\Omegam$). It is possible to increase $\Neff$ and $H_0$ simultaneously while keeping these two redshifts (and both $\Omegam, \Omega_\mathrm{r})$ fixed: thus we expect BAO data to be compatible with arbitrary values of $H_0$ when $\Neff$ fluctuates. However, the degeneracy between $\Neff$ and $H_0$ is limited by BBN data in two main ways.

First, variations in $\Neff$ change the range of $\omegab$ values preferred by deuterium, and different baryon abundances lead to different values of the comoving sound horizon at the time of baryon decoupling. However, this effect is too small to yield any relevant bounds on $(\Neff, H_0)$ when using only the combination of BAO+deuterium data: we checked that BAO+deuterium are easily compatible with values of $H_0$ even much larger than the SH0ES results, together with a very large $\Neff$\,.

Second, the combination of deuterium and helium abundances excludes values of $\Neff$ very different from three. As such, we expect the combined BAO+deuterium+helium data to provide tight bounds on $\Neff$, and thus on $H_0$. In this case, the final bounds for $H_0$ will be dominated by uncertainties on the primordial helium abundance. The role of BAO data is to provide a constraint that allows to translate helium bounds on $\Neff$ into $H_0$ bounds. Of course, the bounds on the Hubble parameter obtained in this way assume that the value of $\Neff$ is the same at BBN time (relevant for helium bounds) and CMB times (relevant for the sound horizon).  Models in which extra radiation is created only after BBN (like the sterile neutrino model of~\cite{Archidiacono:2016kkh} or the Dark Matter converting into Dark Radiation model of~\cite{Bringmann:2018jpr}) naturally evade combined BAO+deuterium+helium bounds on $H_0$, and can potentially accommodate arbitrary large $H_0$ values as long as they also agree with CMB and LSS data.

\begin{figure}[t!]
	\includegraphics[width=0.45\textwidth]{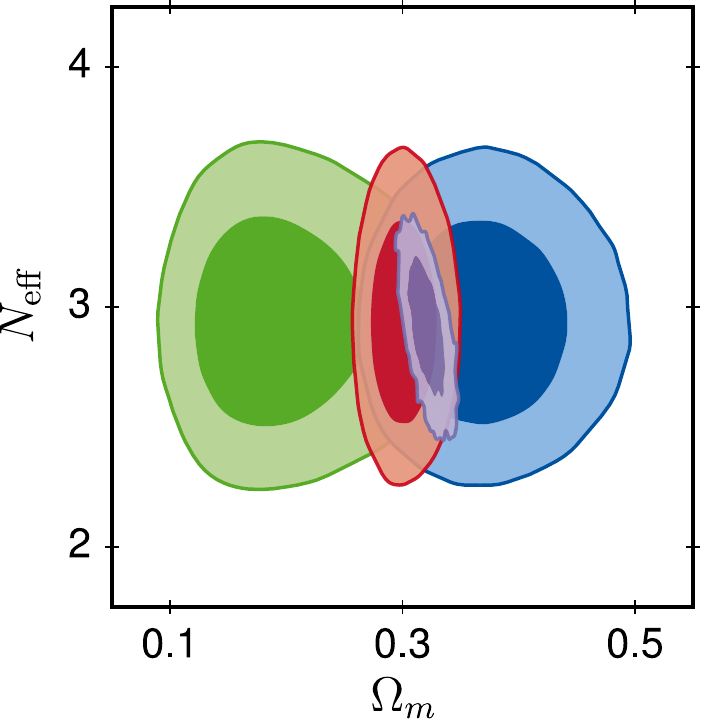}
	\hspace*{0.5em}
	\includegraphics[width=0.46\textwidth]{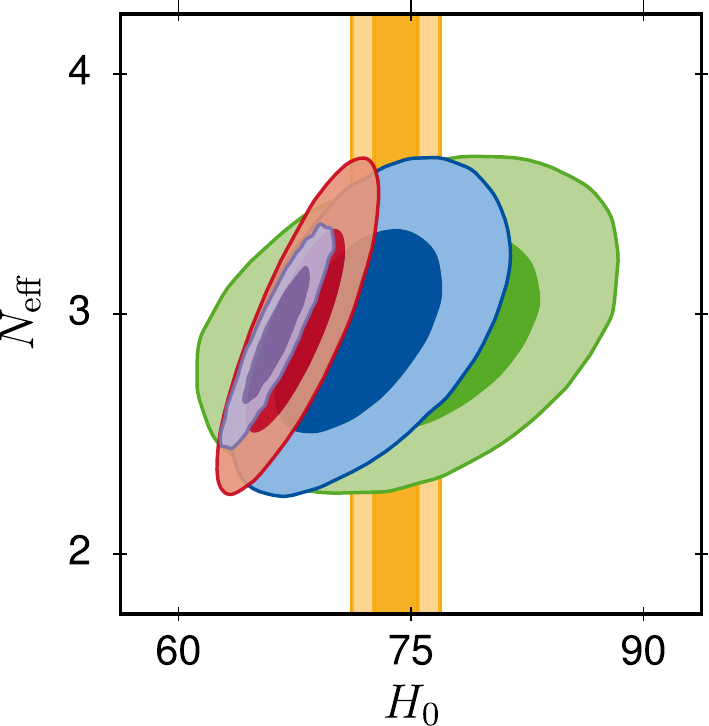}
	\caption{68\% and 95\% confidence levels on $\Omega_m$, $H_0$ and $\Neff$ for the $\Lambda$CDM+$\Neff$ model with various data sets: BAO+BBN for galaxy BAO (blue), Lyman-$\alpha$ BAO (green), and combined (red). Additionally displayed is the Riess et al 2019 measurement (orange), and the Planck 2018 measurement (purple).\label{fig:lyagal}
	}
\end{figure}

We display our results in figure \ref{fig:lyagal}. With our full BAO+BBN data set we measure $H_0 = 67.7 \substack{+2.0 \\ -2.2}$\,km/s/Mpc (68\%C.L.), which is indeed a wider range than with $\Neff=3.046$. It still agrees perfectly with the Planck 2018 measurement $H_0 = 66.4 \pm1.4$\,km/s/Mpc (at the 0.5$\sigma$level). The tension with the SH0ES measurement is reduced, but does not disappear: it only decreases from the 3.2$\sigma$ to the 2.6$\sigma$ level. This result can be attributed to the fact that BBN data alone provide sharp constraints on $\Neff$\,.

\begin{figure}[t!]
	\includegraphics{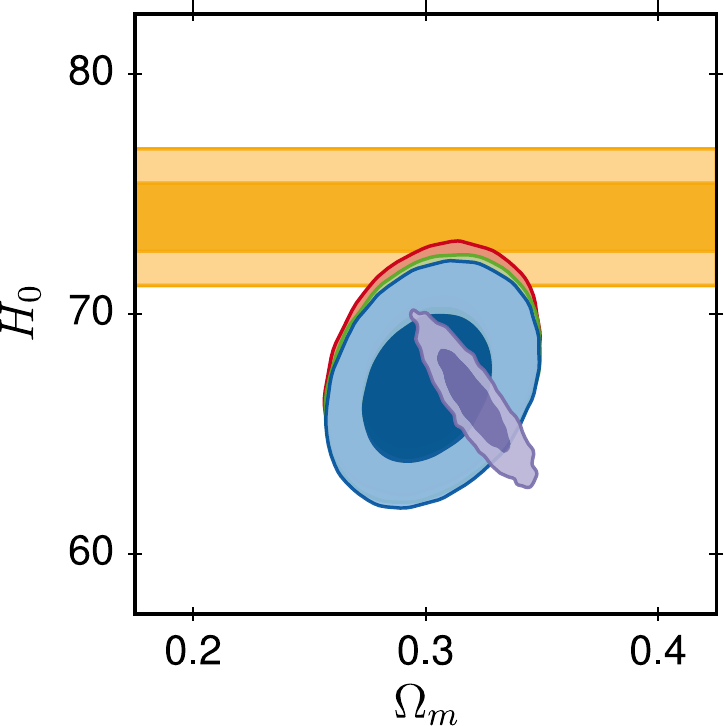}
	\includegraphics{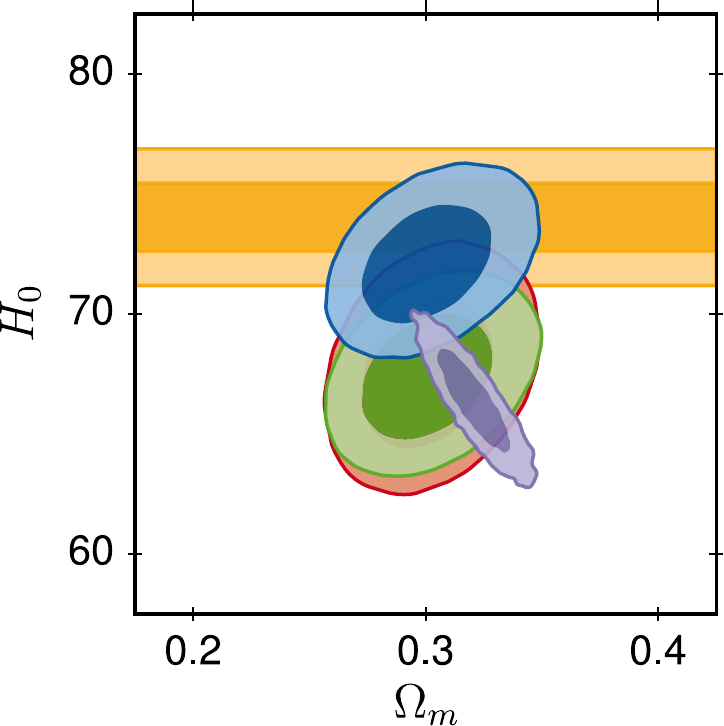}
	\caption{68\% and 95\% confidence levels on $\Omega_m$ and $H_0$ for the $\Lambda$CDM+$\Neff$ model and our combined BAO+BBN data set under various assumptions: \textbf{Left:} BBN predictions taken from {\tt PArthENoPE}-standard (red), {\tt PArthENoPE}-Marcucci (green) and {\tt PRIMAT} (blue). \textbf{Right:} Helium abundance  taken from Aver et al. \cite{Aver:2015iza} (red), Peimbert et al. \cite{Peimbert:2016bdg} (green), and Izotov et al. \cite{Izotov:2014fga} (blue).
	\label{fig:bbn_tests}}
\end{figure}

Again, we test the influence of different assumptions on BBN or helium abundances. The left panel of figure \ref{fig:bbn_tests} shows that the dependence of our results on the three BBN pipelines {\tt PArthENoPE}-standard, {\tt PArthENoPE}-Marcucci, and {\tt PRIMAT} is as small as in the previous case, again because our theoretical error absorbs the impact of different assumptions on nuclear rates. However, the influence of different helium measurements is now very large.  Indeed, the helium measurement directly gives a range of allowed $\Neff$ values. The estimates of Aver et al. \cite{Aver:2015iza} and Peimbert et al. \cite{Peimbert:2016bdg} are close to each other, but Izotov et al. \cite{Izotov:2014fga} find a significantly higher helium abundance, $Y_P=0.2551\pm0.0022$ (68\%C.L.), yielding higher values of $\Neff$ and thus of $H_0$. If that measurement turned out to be the most accurate, BAO+BBN would predict $H_0=72.2 \substack{+1.6\\ -1.7}$\,km/s/Mpc, well compatible with the SH0ES data at the 0.8$\sigma$ level. 

However, the predictions of Izotov et al. \cite{Izotov:2014fga} should be considered with care because they stand far from most previous estimates. The new observations reported in \cite{Izotov:2014fga} are actually included in the compilation of Aver et al. \cite{Aver:2015iza} that we use as a baseline, but the latter authors processed these observations with different assumptions concerning the modeling of several systematic effects, which bring the final result for $Y_P$ in closer agreement with previous estimates.

%% file: Discussion.tex
\newpage
\section{Discussion}
\label{sec:discussion}
References \cite{Addison:2013haa,Aubourg:2014yra,Addison:2017fdm,Blomqvist:2019rah,Cuceu:2019for} have shown that the combination of BAO and BBN measurements can probe the background history of the universe, and provide a measurement of $H_0$ independent of CMB and supernovae. In this paper we have extended these BAO+BBN analyses in several ways. 

First, we have used measurements of both primordial deuterium and helium, employed state-of-the-art BBN codes, and taken into account well-motivated theoretical errors on BBN predictions. With such assumptions, we have been able to confirm the most recent results of references \cite{Blomqvist:2019rah,Cuceu:2019for}, which include new BAO measurement inferred from Lyman-$\alpha$ data in eBOSS DR14; however, we obtain slightly looser bounds on $H_0$. We have shown that in the standard $\Lambda$CDM cosmology, when combining all BAO data in conjunction with BBN, one finds $H_0 = 68.3 \substack{+1.1 \\ -1.2}$\,km/s/Mpc (68\%C.L.), which is in $3.2\sigma$ tension with the SH0ES collaboration measurement, but in very good agreement with the Planck measurement. We remark again that this measurement is \emph{independent} of CMB data.

Second, we have extended the analysis to the $\Lambda$CDM+$\Neff$ model, to see whether the combined BAO+BBN measurements can probe the $H_0 -  \Neff$ degeneracy. When we assume the most conservative and widely accepted estimate of the primordial helium abundance compiled by Aver et al. \cite{Aver:2015iza}, we find $H_0 = 67.7 \substack{+2.0 \\ -2.2}$\,km/s/Mpc (68\%C.L.). Despite the larger uncertainty, the tension with the SH0ES value is only reduced to $2.6\sigma$. We thus find that an enhanced $\Neff$ is not sufficient to accommodate such a large $H_0$ as 74\,km/s/Mpc (68\%C.L.). We find that this conclusion is independent of several possible assumptions on nuclear rates, that mainly affect the relation between $\omegab$ and the deuterium abundance. On the other hand, it does depend significantly on the assumed helium abundance.

We see that the BAO+BBN probe is very useful and informative. It is, of course, subject to statistical and systematic errors that can be improved in the future. Our results depend very strongly on the BAO measurements derived from Lyman-$\alpha$ data, for which the systematic effects require further investigation. On the BBN side, we believe that the part of our analysis related to the deuterium abundance and its relation to $\omegab$ is very robust, since we took into account a conservative theoretical error on deuterium predictions, which propagate uncertainties on the most relevant nuclear rates. We note that the LUNA~\cite{Gustavino:2017veb} experiment will soon provide a direct measurement of the $d(p,\gamma)^3$He rate in the energy range relevant for BBN predictions. This will lead to a reduction in the theoretical error, to an increased precision in the predictions for $\omegab$, and to smaller error bars on $H_0$ from the combined BAO+BBN probe. Finally, in the $\Lambda$CDM+$\Neff$ model, our result depends substantially on the measured primordial helium abundance. This shows that finding a robust experimental consensus on the primordial helium abundance would indirectly help to understand the Hubble tension. Furthermore, a better measurement of the neutrino lifetime would lead to a smaller theoretical error for helium predictions, and thus a smaller uncertainty for both $\Neff$ and $H_0$.

A very interesting and powerful aspect of the BAO+BBN probe is the fact that it is only sensitive to the background cosmology. Thus, our conclusions apply to any model that is degenerate with $\Lambda$CDM+$\Neff$ at the background level. Many models invoked to solve the $H_0$ and $\sigma_8$ tensions involve an enhanced radiation density until the time of photon decoupling, and thus only differ at the perturbation level. Our results show that such models cannot easily solve the $H_0$ tension at the background level: they can accommodate slightly larger values of $H_0$ than in the $\Lambda$CDM case, but they cannot reach $H_0 \simeq 74$\,km/s/Mpc without raising a tension with BAO+BBN data.
However, our bounds do assume that $\Neff$ is the same at BBN and CMB times. Models in which the extra radiation is created after BBN (like e.g. in references~\cite{Archidiacono:2016kkh,Poulin:2016nat,Binder:2017lkj,Bringmann:2018jpr,Poulin:2018cxd}) naturally evade our combined BAO+deuterium+helium bounds.

Finally, we should stress that {\it unlike} the BAO+BBN contours, the CMB contours shown in our figures depend on assumptions at the level of cosmological perturbations. Thus, the CMB contours in figure~\ref{fig:lyagal} only apply to the plain $\Lambda$CDM+$\Neff$ model with extra free-streaming species. For more sophisticated models with e.g interactions in the sector of ordinary neutrinos, sterile neutrinos, Dark Matter, or Dark Radiation, or models with early Dark Energy playing the role of extra radiation, the CMB contours would move around. This is actually the reason for which, in principle, one could hope to find models compatible with CMB+LSS data and at the same time with $H_0 \simeq 74$\,km/s/Mpc, by playing simultaneously with a high $\Neff$ and subtle effects at the perturbation level. Our analysis rules out this possibility under the assumptions of a time-independent $\Neff$ between BBN and CMB times and of our baseline BAO+BBN data set.

%% file: BAO.tex
\section{What do BAO data really measure?} \label{ap:bao_measure}

The matter correlation function $\zeta(r,z)$ of a given cosmological model is related to its matter power spectrum $P_m(k,z)$ by
\begin{equation}
\zeta(r,z) = \int_0^\infty \frac{dk}{k} \frac{k^3}{2 \pi^2} P_m(k,z) j_0(kr)~.
\end{equation}
BAO experiments measure a comoving length $r_\mathrm{BAO}(z)$, called the BAO scale. It is defined as the scale of the local maximum in $\zeta(r,z) $. It is seen either orthogonally to the line of sight, as an angle, or along the line of sight, as a differential distance in redshift space. Since both measurements are independent, they are usually combined with each other. The BAO scale is slightly affected by non-linear evolution at low redshift, which tends to increase it by around 1\%.

This effect is taken into account when analyzing  BAO data: the non-linear evolution is deconvolved, and the reported bounds apply to the BAO scale predicted by {\it linear theory}. In this section we will assume that this removal of non-linear effects is done successfully and without introducing a systematic bias.

The comoving correlation length $r_\mathrm{BAO}(z)$ is usually taken to be the comoving sound horizon $r_s(z)$. Besides, one usually assumes that baryons have instantaneously decoupled at the so-called baryon drag time $t_D$ (associated to redshift $z_D$), such that $r_s$ experiences no dynamics after $t_D$ and remains equal to $ r_s(z_D)$. Let us take for simplicity the example of BAO measurements orthogonal to the line of sight. Such observations are sensitive to the angle under which the BAOs are observed,
\begin{equation}
\theta(z)
= \frac{r_s(z_D)}{r_A(z)}~,
\label{eq:rs}
\end{equation}
where $r_A(z)$ is the comoving angular distance up to redshift $z$. Similarly, other BAO observations will generally depend on $r_s(z_D)$ and another comoving distance, which we generally denote as $r_X(z)$. Their respective ratio is then used to calculate the likelihood.

The comoving sound horizon at baryon drag time is obtained from the simple integral
\begin{equation}
r_s(z_D) = \int^{\tau_D}_0 \!\! c_s(\tau) d\tau = \int^{t_D}_0 \!\! \frac{d t}{3 a(t) \sqrt{1+R(t)}}~,
\end{equation}
where $\tau$ stands for conformal time, $t$ for proper (cosmological) time, and $R=\frac{3 \rho_b}{4 \rho_\gamma}$ is given solely by the CMB temperature today and the baryon density parameter $\omega_b \equiv \Omega_b h^2$.

The identification of the comoving BAO scale with the comoving sound horizon, $r_\mathrm{BAO}(z)=r_s(z_D)$, is a slight oversimplification. In standard cosmology, even if we ignore non-linear effects (or assume that they are well accounted for by the experimental collaborations), the dynamics of the correlation length depends on two types of interactions: the pressure force in the baryon-photon fluid, and gravitational forces. The sound horizon accounts for the first interactions only. Thus, it is usually assumed that the influence of gravity on the correlation length is negligible compared to the one of pressure. However, it is well-known that gravitational interactions between the baryon-photon fluid and other light species shift CMB peaks, and therefore also the BAO scale. This effect is known as the \tquote{neutrino drag} effect when caused by ordinary free-streaming neutrinos \cite{Bashinsky:2003tk,Hou:2011ec,Lesgourgues:2018ncw}; free-streaming particles move at the speed of light $c$, and \tquote{drag} over-densities in the baryon-photon fluid behind them, which move only at the baryon-photon sound speed $c_s < c$. CMB and BAO peaks are then shifted to slightly larger scales, seen under larger angles. In the $\Lambda$CDM model, such an effect is caused by the three active neutrinos, still relativistic at the time of CMB decoupling. In extended cosmological models, the effect increases with more free-streaming relics, i.e. with $N_\mathrm{eff}$. The impact of extra self-interacting relics is much smaller, since over-densities in such relics also propagate at a sound speed smaller than the speed of light (but still not coinciding with the baryon-photon sound speed at all times). The impact of relics with non-zero mass is also reduced, since their velocity is generally smaller around and after their non-relativistic transition.

The experimental collaborations do not rely on the identification $r_\mathrm{BAO}(z)=r_s(z_D)$, because they directly fit the full correlation function $\zeta(r,z)$ to their data. Using the ratio ${r_s(z_D)}/{r_X(z)}$ as a derived parameter, they provide a central value and an error bar for this quantity. With this approach the effect of neutrino drag in the minimal $\Lambda$CDM model is consistently taken into account.

The approximation $r_\mathrm{BAO}(z)= r_s(z_D)$ appears only at the level of BAO likelihoods in sampling codes like {\sc MontePython} \cite{Audren:2012wb,Brinckmann:2018cvx} or {\sc CosmoMC} \cite{Lewis:2002ah}. In these codes the quantity computed for each model and fitted to the data is ${r_s(z_D)}/{r_X(z)}$, not considering the peak in the correlation function. Thus, this approach neglects variations in the amount of gravitational effects, and in particular the impact of extra neutrino drag when $N_\mathrm{eff}$ is enhanced.

This approximation is, however, valid given the current precision level of BAO data, even in models with an enhanced $N_\mathrm{eff}$. This has been shown in details by Thepsuriya and Lewis~\cite{Thepsuriya:2014zda}. We provide here a short argument confirming the conclusions of~\cite{Thepsuriya:2014zda}. We consider two models: the reference model is a minimal $\Lambda$CDM model with three massless neutrinos and $N_\mathrm{eff}=3.046$; the extended models has $N_\mathrm{eff}=4.046$ and identical values for all other parameters like \{$\omega_b$, $\omega_c$, $H_0$\}. For each model we use {\sc class} \cite{Blas:2011rf} to compute the matter power spectrum, the correlation function $\zeta(r,z)$, and the scale $r_\mathrm{BAO}$ of the local maximum at $z=0$ (the result is independent of the redshift). {\sc class} also computes the sound horizon $r_s(z_D)$ by integrating equation~(\ref{eq:rs}). We find that between the reference and extended models, $r_\mathrm{BAO}$ decreases by 3.319\%, while $r_s(z_D)$ decreases by 3.227\%. The relative difference between these two ratio is only of 0.1\%. This is much smaller than errors on BAO measurements, which are of the order of 0.9\% in the best case (in the SDSS DR12 data). We conclude that using variations of $r_s(z_D)$ as an approximation for variations in $r_\mathrm{BAO}$ is valid in this example and with the precision of current experiments. This test allows us to conclude on the validity of the usual approach for all models with enhanced radiation density. For the case of extra free-streaming  radiation, we have shown that the approximation holds for variations of the order of $|\Delta N_\mathrm{eff}|\simeq 1$, which are the maximum variations allowed by current data. As mentioned before, for extra radiation with non-zero masses or self-interactions, gravitational effects like neutrino drag can \textit{only} be smaller. As such, it is legitimate to continue using the usual simplified likelihoods in {\sc MontePython} or {\sc CosmoMC}.

%% file: ms.bbl
\providecommand{\href}[2]{#2}\begingroup\raggedright\begin{thebibliography}{10}

\bibitem{Aghanim:2018eyx}
{\scshape Planck} collaboration, N.~Aghanim et~al., \emph{{Planck 2018 results.
  VI. Cosmological parameters}},  \href{https://arxiv.org/abs/1807.06209}{{\tt
  1807.06209}}.

\bibitem{Riess:2019cxk}
A.~G. Riess, S.~Casertano, W.~Yuan, L.~M. Macri and D.~Scolnic, \emph{{Large
  Magellanic Cloud Cepheid Standards Provide a 1\% Foundation for the
  Determination of the Hubble Constant and Stronger Evidence for Physics beyond
  $\Lambda$CDM}},
  \href{http://dx.doi.org/10.3847/1538-4357/ab1422}{\emph{Astrophys. J.} {\bf
  876} (2019) 85}, [\href{https://arxiv.org/abs/1903.07603}{{\tt 1903.07603}}].

\bibitem{Wong:2019kwg}
K.~C. Wong et~al., \emph{{H0LiCOW XIII. A 2.4\% measurement of $H_{0}$ from
  lensed quasars: $5.3\sigma$ tension between early and late-Universe probes}},
   \href{https://arxiv.org/abs/1907.04869}{{\tt 1907.04869}}.

\bibitem{Freedman:2019jwv}
W.~L. Freedman et~al., \emph{{The Carnegie-Chicago Hubble Program. VIII. An
  Independent Determination of the Hubble Constant Based on the Tip of the Red
  Giant Branch}},  \href{https://arxiv.org/abs/1907.05922}{{\tt 1907.05922}}.

\bibitem{Joudaki:2019pmv}
S.~Joudaki et~al., \emph{{KiDS+VIKING-450 and DES-Y1 combined: Cosmology with
  cosmic shear}},  \href{https://arxiv.org/abs/1906.09262}{{\tt 1906.09262}}.

\bibitem{Vonlanthen:2010cd}
M.~Vonlanthen, S.~Räsänen and R.~Durrer, \emph{{Model-independent
  cosmological constraints from the CMB}},
  \href{http://dx.doi.org/10.1088/1475-7516/2010/08/023}{\emph{JCAP} {\bf 1008}
  (2010) 023}, [\href{https://arxiv.org/abs/1003.0810}{{\tt 1003.0810}}].

\bibitem{Audren:2012wb}
B.~Audren, J.~Lesgourgues, K.~Benabed and S.~Prunet, \emph{{Conservative
  Constraints on Early Cosmology: an illustration of the Monte Python
  cosmological parameter inference code}},
  \href{http://dx.doi.org/10.1088/1475-7516/2013/02/001}{\emph{JCAP} {\bf 1302}
  (2013) 001}, [\href{https://arxiv.org/abs/1210.7183}{{\tt 1210.7183}}].

\bibitem{Audren:2013nwa}
B.~Audren, \emph{{Separate Constraints on Early and Late Cosmology}},
  \href{http://dx.doi.org/10.1093/mnras/stu1457}{\emph{Mon. Not. Roy. Astron.
  Soc.} {\bf 444} (2014) 827--832},
  [\href{https://arxiv.org/abs/1312.5696}{{\tt 1312.5696}}].

\bibitem{Lancaster:2017ksf}
L.~Lancaster, F.-Y. Cyr-Racine, L.~Knox and Z.~Pan, \emph{{A tale of two modes:
  Neutrino free-streaming in the early universe}},
  \href{http://dx.doi.org/10.1088/1475-7516/2017/07/033}{\emph{JCAP} {\bf 1707}
  (2017) 033}, [\href{https://arxiv.org/abs/1704.06657}{{\tt 1704.06657}}].

\bibitem{Oldengott:2017fhy}
I.~M. Oldengott, T.~Tram, C.~Rampf and Y.~Y.~Y. Wong, \emph{{Interacting
  neutrinos in cosmology: exact description and constraints}},
  \href{http://dx.doi.org/10.1088/1475-7516/2017/11/027}{\emph{JCAP} {\bf 1711}
  (2017) 027}, [\href{https://arxiv.org/abs/1706.02123}{{\tt 1706.02123}}].

\bibitem{DiValentino:2017oaw}
E.~Di~Valentino, C.~Boehm, E.~Hivon and F.~R. Bouchet, \emph{{Reducing the
  $H_0$ and $\sigma_8$ tensions with Dark Matter-neutrino interactions}},
  \href{http://dx.doi.org/10.1103/PhysRevD.97.043513}{\emph{Phys. Rev.} {\bf
  D97} (2018) 043513}, [\href{https://arxiv.org/abs/1710.02559}{{\tt
  1710.02559}}].

\bibitem{Kreisch:2019yzn}
C.~D. Kreisch, F.-Y. Cyr-Racine and O.~Dor\'e, \emph{{The Neutrino Puzzle:
  Anomalies, Interactions, and Cosmological Tensions}},
  \href{https://arxiv.org/abs/1902.00534}{{\tt 1902.00534}}.

\bibitem{Park:2019ibn}
M.~Park, C.~D. Kreisch, J.~Dunkley, B.~Hadzhiyska and F.-Y. Cyr-Racine,
  \emph{{$\Lambda$CDM or self-interacting neutrinos? - how CMB data can tell
  the two models apart}},  \href{https://arxiv.org/abs/1904.02625}{{\tt
  1904.02625}}.

\bibitem{Archidiacono:2016kkh}
M.~Archidiacono, S.~Gariazzo, C.~Giunti, S.~Hannestad, R.~Hansen, M.~Laveder
  et~al., \emph{{Pseudoscalar?sterile neutrino interactions: reconciling the
  cosmos with neutrino oscillations}},
  \href{http://dx.doi.org/10.1088/1475-7516/2016/08/067}{\emph{JCAP} {\bf 1608}
  (2016) 067}, [\href{https://arxiv.org/abs/1606.07673}{{\tt 1606.07673}}].

\bibitem{Lesgourgues:2015wza}
J.~Lesgourgues, G.~Marques-Tavares and M.~Schmaltz, \emph{{Evidence for dark
  matter interactions in cosmological precision data?}},
  \href{http://dx.doi.org/10.1088/1475-7516/2016/02/037}{\emph{JCAP} {\bf 1602}
  (2016) 037}, [\href{https://arxiv.org/abs/1507.04351}{{\tt 1507.04351}}].

\bibitem{Buen-Abad:2017gxg}
M.~A. Buen-Abad, M.~Schmaltz, J.~Lesgourgues and T.~Brinckmann,
  \emph{{Interacting Dark Sector and Precision Cosmology}},
  \href{http://dx.doi.org/10.1088/1475-7516/2018/01/008}{\emph{JCAP} {\bf 1801}
  (2018) 008}, [\href{https://arxiv.org/abs/1708.09406}{{\tt 1708.09406}}].

\bibitem{Archidiacono:2019wdp}
M.~Archidiacono, D.~C. Hooper, R.~Murgia, S.~Bohr, J.~Lesgourgues and M.~Viel,
  \emph{{Constraining Dark Matter -- Dark Radiation interactions with CMB, BAO,
  and Lyman-$\alpha$}},  \href{https://arxiv.org/abs/1907.01496}{{\tt
  1907.01496}}.

\bibitem{Poulin:2016nat}
V.~Poulin, P.~D. Serpico and J.~Lesgourgues, \emph{{A fresh look at linear
  cosmological constraints on a decaying dark matter component}},
  \href{http://dx.doi.org/10.1088/1475-7516/2016/08/036}{\emph{JCAP} {\bf 1608}
  (2016) 036}, [\href{https://arxiv.org/abs/1606.02073}{{\tt 1606.02073}}].

\bibitem{Binder:2017lkj}
T.~Binder, M.~Gustafsson, A.~Kamada, S.~M.~R. Sandner and M.~Wiesner,
  \emph{{Reannihilation of self-interacting dark matter}},
  \href{http://dx.doi.org/10.1103/PhysRevD.97.123004}{\emph{Phys. Rev.} {\bf
  D97} (2018) 123004}, [\href{https://arxiv.org/abs/1712.01246}{{\tt
  1712.01246}}].

\bibitem{Bringmann:2018jpr}
T.~Bringmann, F.~Kahlhoefer, K.~Schmidt-Hoberg and P.~Walia, \emph{{Converting
  nonrelativistic dark matter to radiation}},
  \href{http://dx.doi.org/10.1103/PhysRevD.98.023543}{\emph{Phys. Rev.} {\bf
  D98} (2018) 023543}, [\href{https://arxiv.org/abs/1803.03644}{{\tt
  1803.03644}}].

\bibitem{Hooper:2019gtx}
D.~Hooper, G.~Krnjaic and S.~D. McDermott, \emph{{Dark Radiation and Superheavy
  Dark Matter from Black Hole Domination}},
  \href{https://arxiv.org/abs/1905.01301}{{\tt 1905.01301}}.

\bibitem{Poulin:2018cxd}
V.~Poulin, T.~L. Smith, T.~Karwal and M.~Kamionkowski, \emph{{Early Dark Energy
  Can Resolve The Hubble Tension}},
  \href{http://dx.doi.org/10.1103/PhysRevLett.122.221301}{\emph{Phys. Rev.
  Lett.} {\bf 122} (2019) 221301},
  [\href{https://arxiv.org/abs/1811.04083}{{\tt 1811.04083}}].

\bibitem{Agrawal:2019lmo}
P.~Agrawal, F.-Y. Cyr-Racine, D.~Pinner and L.~Randall, \emph{{Rock 'n' Roll
  Solutions to the Hubble Tension}},
  \href{https://arxiv.org/abs/1904.01016}{{\tt 1904.01016}}.

\bibitem{Lin:2019qug}
M.-X. Lin, G.~Benevento, W.~Hu and M.~Raveri, \emph{{Acoustic Dark Energy:
  Potential Conversion of the Hubble Tension}},
  \href{https://arxiv.org/abs/1905.12618}{{\tt 1905.12618}}.

\bibitem{Pan:2019gop}
S.~Pan, W.~Yang, E.~Di~Valentino, E.~N. Saridakis and S.~Chakraborty,
  \emph{{Interacting scenarios with dynamical dark energy: observational
  constraints and alleviation of the $H_0$ tension}},
  \href{https://arxiv.org/abs/1907.07540}{{\tt 1907.07540}}.

\bibitem{Desmond:2019ygn}
H.~Desmond, B.~Jain and J.~Sakstein, \emph{{A local resolution of the Hubble
  tension: The impact of screened fifth forces on the cosmic distance ladder}},
   \href{https://arxiv.org/abs/1907.03778}{{\tt 1907.03778}}.

\bibitem{Addison:2013haa}
G.~E. Addison, G.~Hinshaw and M.~Halpern, \emph{{Cosmological constraints from
  baryon acoustic oscillations and clustering of large-scale structure}},
  \href{http://dx.doi.org/10.1093/mnras/stt1687}{\emph{Mon. Not. Roy. Astron.
  Soc.} {\bf 436} (2013) 1674--1683},
  [\href{https://arxiv.org/abs/1304.6984}{{\tt 1304.6984}}].

\bibitem{Aubourg:2014yra}
E.~Aubourg et~al., \emph{{Cosmological implications of baryon acoustic
  oscillation measurements}},
  \href{http://dx.doi.org/10.1103/PhysRevD.92.123516}{\emph{Phys. Rev.} {\bf
  D92} (2015) 123516}, [\href{https://arxiv.org/abs/1411.1074}{{\tt
  1411.1074}}].

\bibitem{Addison:2017fdm}
G.~E. Addison, D.~J. Watts, C.~L. Bennett, M.~Halpern, G.~Hinshaw and J.~L.
  Weiland, \emph{{Elucidating $\Lambda$CDM: Impact of Baryon Acoustic
  Oscillation Measurements on the Hubble Constant Discrepancy}},
  \href{http://dx.doi.org/10.3847/1538-4357/aaa1ed}{\emph{Astrophys. J.} {\bf
  853} (2018) 119}, [\href{https://arxiv.org/abs/1707.06547}{{\tt
  1707.06547}}].

\bibitem{Blomqvist:2019rah}
M.~Blomqvist et~al., \emph{{Baryon acoustic oscillations from the
  cross-correlation of Ly$\alpha$ absorption and quasars in eBOSS DR14}},
  \href{https://arxiv.org/abs/1904.03430}{{\tt 1904.03430}}.

\bibitem{Cuceu:2019for}
A.~Cuceu, J.~Farr, P.~Lemos and A.~Font-Ribera, \emph{{Baryon Acoustic
  Oscillations and the Hubble Constant: Past, Present and Future}},
  \href{https://arxiv.org/abs/1906.11628}{{\tt 1906.11628}}.

\bibitem{Agathe:2019vsu}
V.~de~Sainte~Agathe et~al., \emph{{Baryon acoustic oscillations at $z = 2.34$
  from the correlations of Ly$\alpha$ absorption in eBOSS DR14}},
  \href{https://arxiv.org/abs/1904.03400}{{\tt 1904.03400}}.

\bibitem{Bashinsky:2003tk}
S.~Bashinsky and U.~Seljak, \emph{{Neutrino perturbations in CMB anisotropy and
  matter clustering}},
  \href{http://dx.doi.org/10.1103/PhysRevD.69.083002}{\emph{Phys. Rev.} {\bf
  D69} (2004) 083002}, [\href{https://arxiv.org/abs/astro-ph/0310198}{{\tt
  astro-ph/0310198}}].

\bibitem{Hou:2011ec}
Z.~Hou, R.~Keisler, L.~Knox, M.~Millea and C.~Reichardt, \emph{{How Massless
  Neutrinos Affect the Cosmic Microwave Background Damping Tail}},
  \href{http://dx.doi.org/10.1103/PhysRevD.87.083008}{\emph{Phys. Rev.} {\bf
  D87} (2013) 083008}, [\href{https://arxiv.org/abs/1104.2333}{{\tt
  1104.2333}}].

\bibitem{Lesgourgues:2018ncw}
J.~Lesgourgues, G.~Mangano, G.~Miele and S.~Pastor, \emph{{Neutrino
  Cosmology}}.
\newblock Cambridge University Press, 2013.

\bibitem{Thepsuriya:2014zda}
K.~Thepsuriya and A.~Lewis, \emph{{Accuracy of cosmological parameters using
  the baryon acoustic scale}},
  \href{http://dx.doi.org/10.1088/1475-7516/2015/01/034}{\emph{JCAP} {\bf 1501}
  (2015) 034}, [\href{https://arxiv.org/abs/1409.5066}{{\tt 1409.5066}}].

\bibitem{Eisenstein:1997ik}
D.~J. Eisenstein and W.~Hu, \emph{{Baryonic features in the matter transfer
  function}}, \href{http://dx.doi.org/10.1086/305424}{\emph{Astrophys. J.} {\bf
  496} (1998) 605}, [\href{https://arxiv.org/abs/astro-ph/9709112}{{\tt
  astro-ph/9709112}}].

\bibitem{2011MNRAS.416.3017B}
F.~{Beutler}, C.~{Blake}, M.~{Colless}, D.~H. {Jones}, L.~{Staveley-Smith},
  L.~{Campbell} et~al., \emph{{The 6dF Galaxy Survey: baryon acoustic
  oscillations and the local Hubble constant}},
  \href{http://dx.doi.org/10.1111/j.1365-2966.2011.19250.x}{\emph{MNRAS} {\bf
  416} (Oct, 2011) 3017--3032}, [\href{https://arxiv.org/abs/1106.3366}{{\tt
  1106.3366}}].

\bibitem{Ross:2014qpa}
A.~J. Ross, L.~Samushia, C.~Howlett, W.~J. Percival, A.~Burden and M.~Manera,
  \emph{{The clustering of the SDSS DR7 main Galaxy sample – I. A 4 per cent
  distance measure at $z = 0.15$}},
  \href{http://dx.doi.org/10.1093/mnras/stv154}{\emph{Mon. Not. Roy. Astron.
  Soc.} {\bf 449} (2015) 835--847},
  [\href{https://arxiv.org/abs/1409.3242}{{\tt 1409.3242}}].

\bibitem{Ata:2017dya}
M.~Ata et~al., \emph{{The clustering of the SDSS-IV extended Baryon Oscillation
  Spectroscopic Survey DR14 quasar sample: first measurement of baryon acoustic
  oscillations between redshift 0.8 and 2.2}},
  \href{http://dx.doi.org/10.1093/mnras/stx2630}{\emph{Mon. Not. Roy. Astron.
  Soc.} {\bf 473} (2018) 4773--4794},
  [\href{https://arxiv.org/abs/1705.06373}{{\tt 1705.06373}}].

\bibitem{Alam:2016hwk}
{\scshape BOSS} collaboration, S.~Alam et~al., \emph{{The clustering of
  galaxies in the completed SDSS-III Baryon Oscillation Spectroscopic Survey:
  cosmological analysis of the DR12 galaxy sample}},
  \href{http://dx.doi.org/10.1093/mnras/stx721}{\emph{Mon. Not. Roy. Astron.
  Soc.} {\bf 470} (2017) 2617--2652},
  [\href{https://arxiv.org/abs/1607.03155}{{\tt 1607.03155}}].

\bibitem{Brinckmann:2018cvx}
T.~Brinckmann and J.~Lesgourgues, \emph{{MontePython 3: boosted MCMC sampler
  and other features}},  \href{https://arxiv.org/abs/1804.07261}{{\tt
  1804.07261}}.

\bibitem{Blas:2011rf}
D.~Blas, J.~Lesgourgues and T.~Tram, \emph{{The Cosmic Linear Anisotropy
  Solving System (CLASS) II: Approximation schemes}},
  \href{http://dx.doi.org/10.1088/1475-7516/2011/07/034}{\emph{JCAP} {\bf 1107}
  (2011) 034}, [\href{https://arxiv.org/abs/1104.2933}{{\tt 1104.2933}}].

\bibitem{Lewis:1999bs}
A.~Lewis, A.~Challinor and A.~Lasenby, \emph{{Efficient computation of CMB
  anisotropies in closed FRW models}},
  \href{http://dx.doi.org/10.1086/309179}{\emph{Astrophys. J.} {\bf 538} (2000)
  473--476}, [\href{https://arxiv.org/abs/astro-ph/9911177}{{\tt
  astro-ph/9911177}}].

\bibitem{Cooke:2017cwo}
R.~J. Cooke, M.~Pettini and C.~C. Steidel, \emph{{One Percent Determination of
  the Primordial Deuterium Abundance}},
  \href{http://dx.doi.org/10.3847/1538-4357/aaab53}{\emph{Astrophys. J.} {\bf
  855} (2018) 102}, [\href{https://arxiv.org/abs/1710.11129}{{\tt
  1710.11129}}].

\bibitem{Aver:2015iza}
E.~Aver, K.~A. Olive and E.~D. Skillman, \emph{{The effects of He I
  $\lambda$10830 on helium abundance determinations}},
  \href{http://dx.doi.org/10.1088/1475-7516/2015/07/011}{\emph{JCAP} {\bf 1507}
  (2015) 011}, [\href{https://arxiv.org/abs/1503.08146}{{\tt 1503.08146}}].

\bibitem{Peimbert:2016bdg}
A.~Peimbert, M.~Peimbert and V.~Luridiana, \emph{{The primordial helium
  abundance and the number of neutrino families}}, {\emph{Rev. Mex. Astron.
  Astrofis.} {\bf 52} (2016) 419},
  [\href{https://arxiv.org/abs/1608.02062}{{\tt 1608.02062}}].

\bibitem{Izotov:2014fga}
Y.~I. Izotov, T.~X. Thuan and N.~G. Guseva, \emph{{A new determination of the
  primordial He abundance using the He i $\lambda$10830 Å emission line:
  cosmological implications}},
  \href{http://dx.doi.org/10.1093/mnras/stu1771}{\emph{Mon. Not. Roy. Astron.
  Soc.} {\bf 445} (2014) 778--793},
  [\href{https://arxiv.org/abs/1408.6953}{{\tt 1408.6953}}].

\bibitem{Consiglio:2017pot}
R.~Consiglio, P.~F. de~Salas, G.~Mangano, G.~Miele, S.~Pastor and O.~Pisanti,
  \emph{{PArthENoPE reloaded}},
  \href{http://dx.doi.org/10.1016/j.cpc.2018.06.022}{\emph{Comput. Phys.
  Commun.} {\bf 233} (2018) 237--242},
  [\href{https://arxiv.org/abs/1712.04378}{{\tt 1712.04378}}].

\bibitem{Adelberger:2010qa}
E.~G. Adelberger et~al., \emph{{Solar fusion cross sections II: the pp chain
  and CNO cycles}},
  \href{http://dx.doi.org/10.1103/RevModPhys.83.195}{\emph{Rev. Mod. Phys.}
  {\bf 83} (2011) 195}, [\href{https://arxiv.org/abs/1004.2318}{{\tt
  1004.2318}}].

\bibitem{Marcucci:2015yla}
L.~E. Marcucci, G.~Mangano, A.~Kievsky and M.~Viviani, \emph{{Implication of
  the proton-deuteron radiative capture for Big Bang Nucleosynthesis}},
  \href{http://dx.doi.org/10.1103/PhysRevLett.116.102501,
  10.1103/PhysRevLett.117.049901}{\emph{Phys. Rev. Lett.} {\bf 116} (2016)
  102501}, [\href{https://arxiv.org/abs/1510.07877}{{\tt 1510.07877}}].

\bibitem{Pitrou:2018cgg}
C.~Pitrou, A.~Coc, J.-P. Uzan and E.~Vangioni, \emph{{Precision big bang
  nucleosynthesis with improved Helium-4 predictions}},
  \href{http://dx.doi.org/10.1016/j.physrep.2018.04.005}{\emph{Phys. Rept.}
  {\bf 754} (2018) 1--66}, [\href{https://arxiv.org/abs/1801.08023}{{\tt
  1801.08023}}].

\bibitem{Nollett:2011aa}
K.~M. Nollett and G.~P. Holder, \emph{{An analysis of constraints on
  relativistic species from primordial nucleosynthesis and the cosmic microwave
  background}},  \href{https://arxiv.org/abs/1112.2683}{{\tt 1112.2683}}.

\bibitem{Gustavino:2017veb}
C.~Gustavino, \emph{{Underground Study of Big Bang Nucleosynthesis in the
  Precision Era of Cosmology}},
  \href{http://dx.doi.org/10.1051/epjconf/201713601009}{\emph{EPJ Web Conf.}
  {\bf 136} (2017) 01009}.

\bibitem{Lewis:2002ah}
A.~Lewis and S.~Bridle, \emph{{Cosmological parameters from CMB and other data:
  A Monte Carlo approach}},
  \href{http://dx.doi.org/10.1103/PhysRevD.66.103511}{\emph{Phys. Rev.} {\bf
  D66} (2002) 103511}, [\href{https://arxiv.org/abs/astro-ph/0205436}{{\tt
  astro-ph/0205436}}].

\end{thebibliography}\endgroup
